\documentstyle[12pt,psfig,axodraw,a4]{article} 
\def\bild#1#2{    
        \vspace*{-5mm}
        \begin{center}
        \begin{math}
        \epsfxsize#2cm
        \epsffile{#1}
        \end{math}
        \end{center}
        }
\newcommand{\vs}{\vspace{-0.25cm}}

\begin{document} 


\begin{center}
\Large{\bf
Role of Chiral Symmetry in the \\Nuclear Many-Body Problem}\footnote{Invited
talk presented by W. Weise at the NATO Advanced Research Workshop: The Nuclear 
Many-Body Problem 2001, 2.-5.6., Brijuni, Croatia}\footnote{Work 
supported in part by BMBF and DFG}  

\bigskip 

\bigskip

N. Kaiser$\,^a$ and W. Weise$\,^{a,b}$\\

\bigskip

{\small
$^a$\,Physik Department, Technische Universit\"{a}t M\"{u}nchen, D-85747
Garching, Germany\\

$^b$\, ECT$^*$, I-38050 Villazzano (Trento), Italy}

\smallskip
\end{center}

\bigskip

\begin{abstract}
The role of chiral (pion) dynamics in nuclear matter is
reviewed. Contributions to the energy per particle from one- and two-pion
exchange are calculated systematically, and it is demonstrated that already at
order $k^4_f$ in the Fermi momentum, two-pion exchange produces realistic
nuclear binding together with very reasonable values for the compressibility
and the asymmetry energy. Further implications of these results are discussed.
\end{abstract}

\bigskip
\bigskip

\section{Introduction}

The present status of the nuclear matter problem is that a quantitatively
successful description can be achieved, using advanced many-body techniques
\cite{1}, in a non-relativistic framework when invoking an adjustable
three-body force. Alternative relativistic mean field approaches, including
non-linear terms with adjustable parameters, are also widely used for the
calculation of nuclear matter properties and finite nuclei \cite{2}. At a more
basic level, the Dirac-Brueckner method \cite{3} solves a relativistically
improved Bethe-Goldstone equation with one-boson exchange $NN$-interactions.

In recent years a novel approach to the $NN$-interaction based in effective
field theory (in particular, chiral perturbation theory) has emerged
\cite{4,5}. The key element is a power counting scheme which separates long-
and short-distance dynamics. Methods of effective field theory  have also been
applied to systems of finite density \cite{6}.

The purpose of this presentation is to point out the importance of explicit
pion dynamics in the nuclear many-body problem. While pion exchange processes
are well established as generators of the long and intermediate range
$NN$-interaction, their role in nuclear matter is less evident. The one-pion
exchange Hartree term vanishes identically, and the leading
Fock exchange term is small. Two-pion exchange mechanisms are commonly hidden 
behind a purely
phenomenological scalar ("sigma"-) mean field which is fitted to empirical data
but has no basic justification. This is an unsatisfactory situation which calls
for a deeper understanding. We report on steps and thoughts in this direction,
following ref. \cite{7}. Our approach is closely related to the work of Lutz et
al. in ref. \cite{6}.

Before passing on to our calculation it is useful to draw attention to the
following fact. A simple but realistic parametrization of the energy per
particle, $\bar{E}(k_f) = E/A$, of isospin symmetric nuclear matter is given in
powers of the Fermi momentum $k_f$ as
\begin{equation}
\bar{E} (k_f) = \frac{3 k^2_f}{10 \, M} - \alpha \frac{k^3_f}{M^2} + \beta
\frac{k^4_f}{M^3} ,
\end{equation}
where the nucleon density is $\rho = 2 k^3_f / 3 \pi^2$ as usual, and $M =
0.939 \, GeV$ is the free nucleon mass. The first term is the kinetic energy of
a Fermi gas. Adjusting the (dimensionless) parameters $\alpha$ and $\beta$ to
the equilibrium density, $\rho_0 = 0.16 \, fm^{-3} \, (k_{f0} = 1.33 \,
fm^{-1})$ and $\bar{E}_0 = \bar{E} (k_{f 0}) = -16 \, MeV$, gives $\alpha
= 5.27$ and $\beta = 12.22$. The compression modulus $K = k_{f 0}^2
(\partial^2 \bar{E} (k_f) / \partial k^2_f)_{k_{f0}}$ is then predicted at $K
= 236 \, MeV$, well in line with empirically deduced values, and the density
dependence of $\bar{E}(k_f)$ using eq. (1) is remarkably close to the one 
resulting from the realistic many-body calculations of the Urbana group
\cite{8}. 
 
\section{Chiral in-medium perturbation theory}
The tool to investigate the implications of spontaneous and explicit chiral
symmetry breaking in QCD is chiral perturbation theory. Observables are
calculated within the framework of an effective field theory of Goldstone
bosons (pions) interacting with the lowest-mass baryons (nucleons). The
diagrammatic expansion of this low-energy theory in the number of loops has a
one-to-one correspondence to a systematic expansion of observables in small
external momenta and the pion (or quark) mass.

In nuclear matter, the relevant momentum scale is the Fermi momentum $k_f$. At
the empirical saturation point, $k_{f 0} \simeq 2m_{\pi}$, so the Fermi
momentum and the pion mass are of comparable magnitude at the densities of 
interest. This
immediately implies that pions {\it must} be included as {\it explicit} degrees
of freedom: their propagation in matter is relevant. Pionic effects cannot be
accounted for simply by adjusting coefficients of local $NN$ contact
interactions.

Both $k_f$ and $m_{\pi}$ are small compared to the characteristic
chiral scale, $4 \pi f_{\pi} \simeq 1.2 \, GeV$, which involves the pion decay
constant $f_{\pi} = 0.092 \, GeV$. Consequently, the equation of state of
nuclear matter as given by chiral perturbation theory will be represented as an
expansion in powers of the Fermi momentum. The expansion coefficients are
non-trivial functions of $k_f/ m_{\pi}$, the dimensionless ratio of the two
relevant scales inherent to the problem.

The chiral effective Lagrangian generates the basic pion-nucleon coupling 
terms: the
Tomo- zawa-Weinberg $\pi \pi NN$ contact vertex, $(1/4 f^2_{\pi}) (q^{\mu}_b -
q^{\mu}_a) \gamma_{\mu} \epsilon_{abc} \tau_c$, and the pseudovector $\pi NN$
vertex, $(g_A / 2 f_{\pi}) q^{\mu}_a \gamma_{\mu} \gamma_5 \tau_a$, where
$q_{a,b}$ denotes (outgoing) pion four-momenta and $g_A$ is the axial vector
coupling constant (we choose $g_A = 1.3$ so that the Goldberger-Treiman 
relation
$g_{\pi N} = g_A M /f_{\pi}$ gives the empirical $\pi N$ coupling constant,
$g_{\pi N} = 13.2$).

The only new ingredient in performing calculations at
finite density (as compared to evaluations of scattering processes in vacuum)
is the in-medium nucleon propagator. For a relativistic nucleon with
four-momentum $p^{\mu} = (p_0, \vec{p} \,)$ it reads
\begin{equation}
(\not \!p + M) \left\{ \frac{i}{p^2 - M^2 + i \varepsilon} - 2 \pi \delta (p^2
- M^2) \theta (p_0) \theta(k_f - | \vec{p}\, |) \right\}.
\end{equation}
The second term is the medium insertion which accounts for the fact that the
ground state of the system has changed from an "empty" vacuum to a filled Fermi
sea of nucleons. Diagrams can then be organized systematically  in the number
of medium insertions, and an expansion is performed in leading inverse powers
of the nucleon mass, consistently with the $k_f$-expansion.

Our "inward-bound" strategy \cite{7} is now as follows. One starts at large
distances (small $k_f$) and systematically generates the pion-induced
correlations between nucleons as they develop with decreasing distance
(increasing $k_f$). The present calculations are performed to 3-loop order
(including terms up to order $k^5_f$) and incorporate one- and two-pion
exchange processes. The procedure involves one single momentum space cutoff
$\Lambda$ which encodes dynamics at short distances not resolved explicitly in
the effective low-energy theory. This cutoff scale $\Lambda$ is the only free
parameter which has to be fine-tuned. (Alternatively, and equivalently, one
could use dimensional regularization and introduce short-distance physics
through adjustable $NN$ contact terms).

We now outline the leading contributions to the energy per particle $\bar{E}
(k_f)$. The kinetic energy including first order relativistic corrections is
\begin{equation}
\bar{E}_{kin} (k_f) = \frac{3 k^2_f}{10 M} \left( 1 - \frac{5 k^2_f}{28 M^2} 
\right).
\end{equation}
Terms of order $k^6_f$ are already negligibly small. At least from this
perspective, nuclear matter is a non-relativistic system.

Nuclear chiral dynamics up to three-loop order introduces the diagrams,
Fig. 1. They include the one-pion exchange (OPE) Fock term, iterated OPE and
irreducible two-pion exchange.

\bigskip

\bild{brijuni1.epsi}{12}
\bild{brijuni2.epsi}{12}

{\it Fig.\,1: In-medium chiral perturbation theory: One-pion exchange Fock term
(upper left), iterated one-pion exchange (upper middle and right) and examples
of irreducible two-pion exchange terms. See ref. \cite{7} for details.}

\bigskip

Medium insertions are systematically applied on all nucleon propagators, and
the relevant loop integrations yield results which can be written in analytic
form for all pieces.

The OPE Fock term becomes
\begin{equation}
\bar{E}_{1 \pi} (k_f) = \frac{g^2_A m^3_{\pi}}{(4 \pi f_{\pi})^2} \left[ F 
\left(
\frac{k_f}{m_{\pi}} \right) + \frac{m^2_{\pi}}{M^2} G \left(
\frac{k_f}{m_{\pi}} \right) \right],
\end{equation}
where $F$ and $G$ are functions of the dimensionless variable $k_f /
m_{\pi}$. They are given explicitly in ref. \cite{7}. All finite parts of
iterated OPE and irreducible two-pion exchange are of the generic form
\begin{equation}
\bar{E}_{2 \pi} (k_f) = \frac{m^4_{\pi}}{(4 \pi f_{\pi})^4} \left[ g^4_A M H_4
\left( \frac{k_f}{m_{\pi}} \right) + m_{\pi} H_5 \left( \frac{k_f}{m_{\pi}}
\right) \right],
\end{equation}
with the functions $H_{4,5}$ again given explicitly in ref. \cite{7}. All power
divergences specific to cutoff regularization are summarized in the expression
\begin{equation}
\bar{E}_{\Lambda} (k_f) = \frac{\Lambda k^3_f}{(4 \pi f_{\pi})^4} [ - 10 g^4_A
M + (3 g^2_A + 1) (g^2_A - 1) \Lambda ],
\end{equation}
where the attractive and dominant first term in the brackets arises from
iterated OPE. Note that this term could have been generated, equivalently, by a
$NN$ contact interaction with appropriate coupling strength.

\section{Results}
\subsection{Nuclear matter equation of state}
A striking feature of the chiral dynamics approach is the simplicity of the
saturation mechanism for isospin-symmetric nuclear matter. Before turning to
the presentation of detailed results, it is instructive first to discuss the
situation in the exact chiral limit, $m_{\pi} = 0$. The basic saturation
mechanism can already be demonstrated by truncating the one- and two-pion
exchange diagrams at order $k^4_f$. We can make straightforward contact with
the parametrization (1) of the energy per particle and identify the
coefficients $\alpha$ and $\beta$ of the $k^3_f$ and $k^4_f$ terms,
respectively. The result for $\alpha$ in the chiral limit is:
\begin{equation}
\alpha = \frac{10 \Lambda}{M} \left( \frac{g_{\pi N}}{4 \pi} \right)^4 - \left(
\frac{g_{\pi N}}{4 \pi}
\right)^2 ,
\end{equation}
where we have neglected the small correction proportional to $\Lambda^2$ in
eq. (6). The strongly attractive leading term in eq. (7) is accompanied by the
(weakly repulsive) one-pion exchange Fock term.

The $k^3_f$-contribution to $\bar{E} (k_f)$ would lead to collapse of the 
many-body
system. The stabilizing $k^4_f$-term is controlled by the coefficient
(calculated again in the chiral limit)
\begin{equation}
\beta = \frac{3}{70} \left( \frac{g_{\pi N}}{4 \pi} \right)^4 ( 4 \pi^2 + 237 -
24 \ln 2) - \frac{3}{56} = 13.55,
\end{equation}
a unique and parameterfree result to this order. Here the two-pion exchange
dynamics produces repulsion of just the right magnitude to achieve saturation:
the result, eq. (8), is within $10 \%$ of the empirical $\beta =
12.2$. Adjustment of the short-distance scale $\Lambda$ between $0.5$ and $0.6 
\, GeV$
easily leads to a stable minimum of $\bar{E} (k_f)$ in the proper range of
density and binding energy.

The full 3-loop  chiral dynamics result for $\bar{E} (k_f)$ in symmetric
nuclear matter, using $m_{\pi} = 135 \, MeV$ (the neutral pion mass), is shown
in Fig. 2 together with a realistic many-body calculation. The outcome is
remarkable: with one single parameter $\Lambda = 0.65 \, GeV$ fixed to the
            value $\bar{E}_0 = -15.3 \, MeV$ at equilibrium, perturbative pion
            dynamics alone produces an equation of state which follows that of
            much more sophisticated calculations up to about three times the
            density of nuclear matter. The predicted compression modulus is $K
            = 255 \, MeV$, well in line with the "empirical" $K = (250 \pm 25)
            \, MeV$ deduced in refs. \cite{9,10}.

\bigskip

\bild{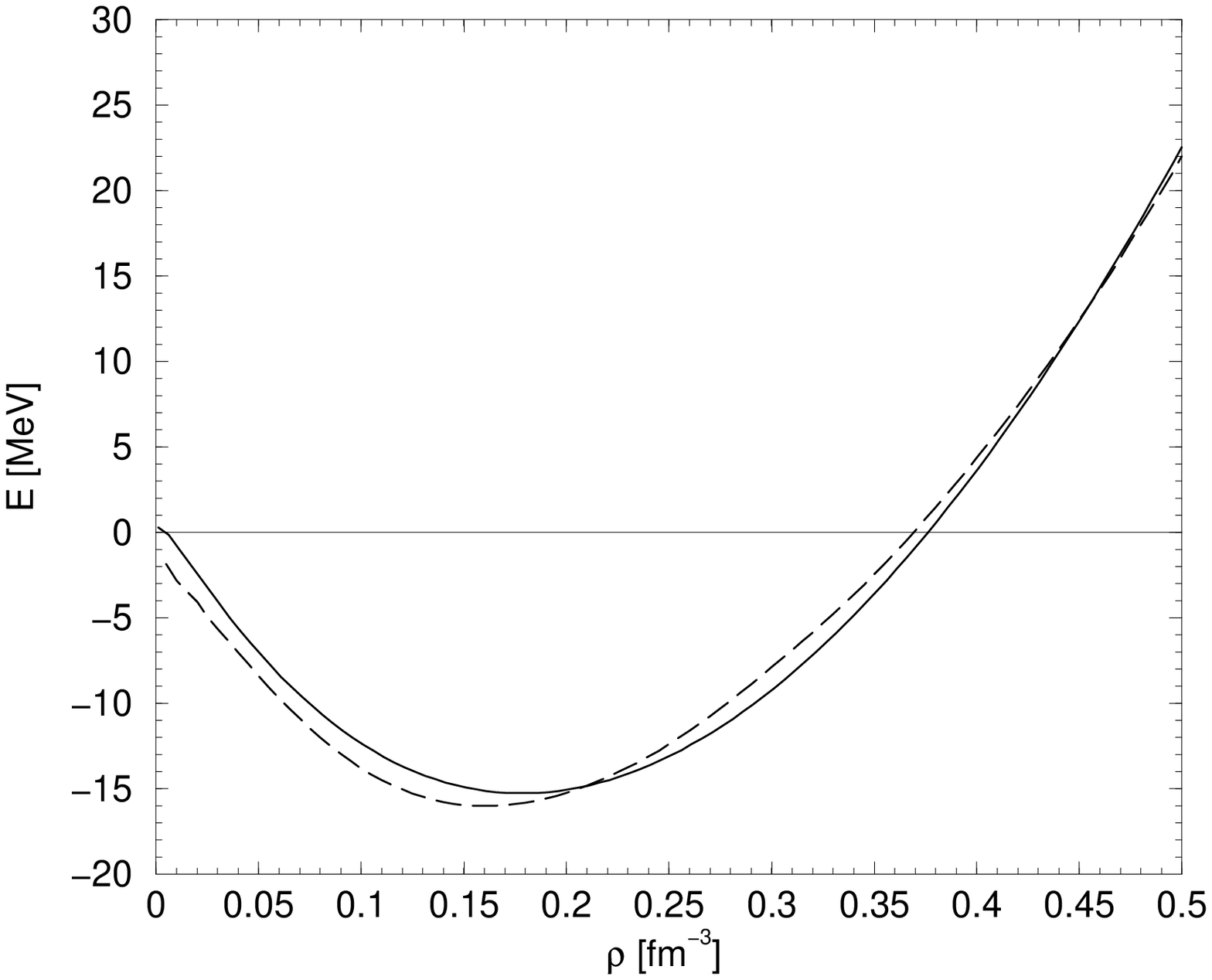}{11.5}

{\it Fig.\,2: Energy per particle, $\bar{E} (k_f)$, of symmetric nuclear matter
derived from chiral one- and two-pion exchange (solid line) \cite{7}. The
cutoff scale is $\Lambda = 646 \, MeV$. The dashed line is the result of ref. 
\cite{8}.}

\bigskip

\bild{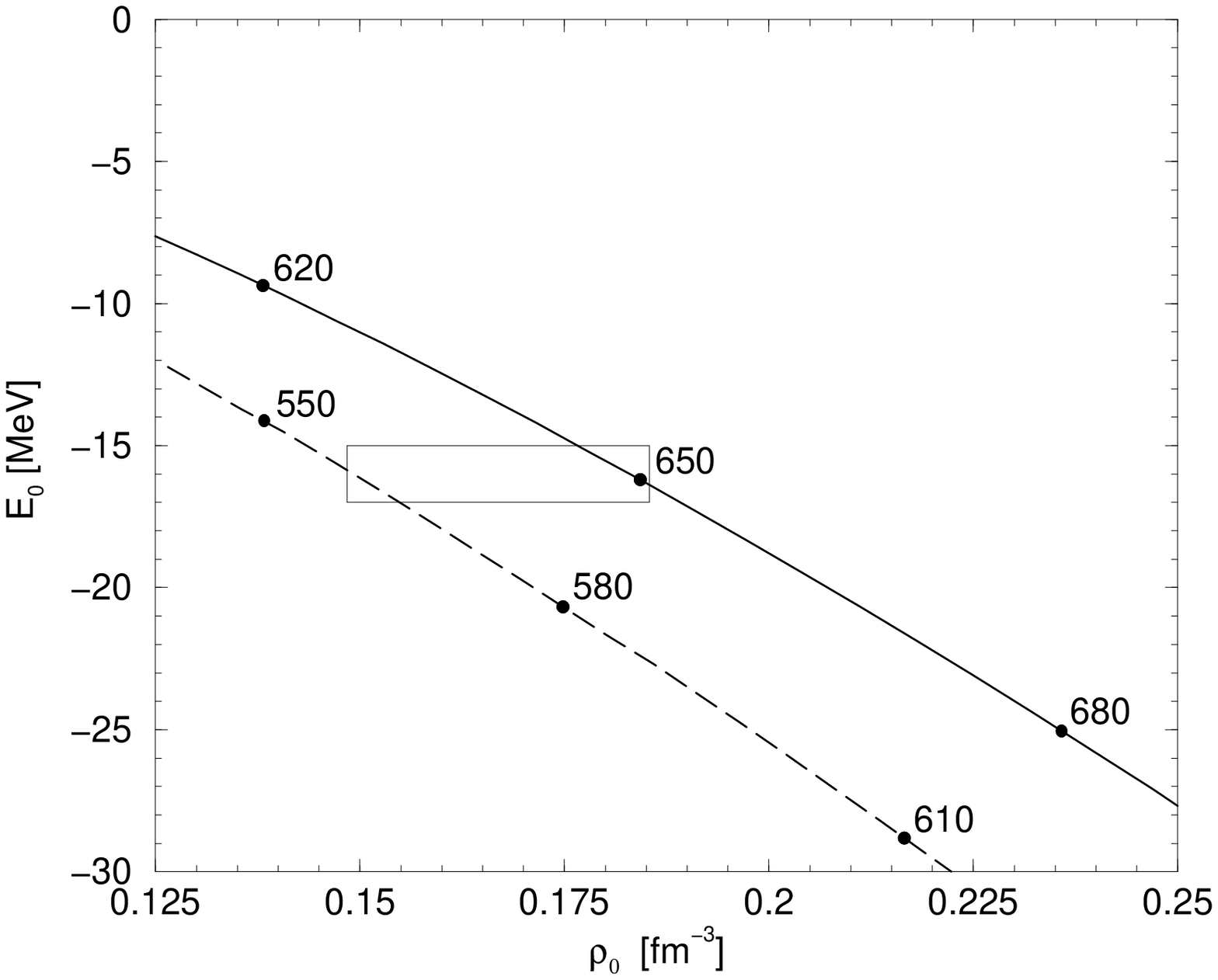}{11.5}

{\it Fig.\,3: Nuclear matter saturation point $(\bar{E}_0, \rho_0)$ at finite
pion  mass $m_{\pi} = 135 \, MeV$ (solid line) and in the chiral limit $m_{\pi}
= 0$  (dashed line) as function of the cutoff scale $\Lambda$ (given in
$MeV$). The inserted rectangle corresponds to the empirical saturation point
including its uncertainties.}

\bigskip

It is interesting to examine the variation with the pion mass in this
context. Fig. 2 displays the position of the nuclear matter saturation point
$(\bar{E}_0, \rho_0)$, first in the chiral limit and then using the physical
$m_{\pi}$, along lines with varying short-distance scale parameter
$\Lambda$. Evidently, explicit chiral symmetry breaking by $m_{\pi}$ is not a
qualitatively decisive feature for saturation; it influences, however, the
quantitative fine-tuning of $\Lambda$.

\subsection{Asymmetry energy}
The specific isospin dependence of two-pion exchange should have its distinct
influence on the behavior of asymmetric nuclear matter, with increasing excess
of neutrons over protons. We introduce as usual the asymmetry parameter
$\delta = (\rho_n - \rho_p) / \rho = (N-Z) / (N+Z)$, keeping the total density
$\rho = \rho_n + \rho_p = 2 k^3_f/ 3 \pi^2$ constant. The proton and neutron
densities are $\rho_{p,n} = k^3_{p,n} / 3 \pi^2$ in terms of the corresponding
Fermi momenta. Without change of any input, we have calculated \cite{7} the
asymmetry energy $A (k_f)$ defined by
\begin{equation}
\bar{E}_{as} (k_p , k_n ) = \bar{E} (k_f) + \delta^2 A (k_f) +...
\end{equation}
The result at nuclear matter density is $A_0 = A (k_{f0}) = 33.8 \,
MeV$. This is in very good agreement with the empirical value $A_0 = 33.2 \,
MeV$ derived from extensive fits to nuclide masses \cite{11}.

Extrapolations to higher density work roughly up to $\rho \simeq 1.5 \,
\rho_0$. At still higher densities, there are indications that non-trivial
isospin dependence beyond one- and two-pion exchange starts to play a role. A
similar statement holds for pure neutron matter which is properly unbound, but
its predicted equation of state starts to deviate from that of realistic
many-body calculations at neutron densities larger than $0.2 \, fm^{-3}$.

\subsection{Nuclear mean field from chiral dynamics}
The in-medium three-loop calculation of the energy per particle defines the
(momentum dependent) self-energy of a single nucleon in nuclear matter up to
two-loop order. The real part of the resulting single particle potential in
isospin-symmetric matter at the saturation point, for a nucleon with zero
momentum, comes out as \cite{12}
\begin{equation}
U (p = 0, k_{f0}) = -53.2 \, MeV,
\end{equation}
using exactly the same one- and two-pion exchange input that has led to the
solid curve in Fig. 2. The momentum dependence of $U (p, k_{f0})$ can be 
rephrased in
terms of an average effective nucleon mass $M^* \simeq 0.8 \, M$ at nuclear
matter density, and the imaginary part of the potential for a nucleon-hole at
the bottom of the Fermi sea is predicted to be about $30 \, MeV$. All these
numbers are remarkably close to the empirically deduced ones.

\section{Summary and outlook}
Explicit pion dynamics originating from the spontaneously broken chiral 
symmetry
of QCD is an important aspect of the nuclear many-body problem. In-medium
chiral perturbation theory, with one single cutoff scale $\Lambda \simeq 0.65
\, GeV$ introduced to regularize the few divergent parts associated with
two-pion exchange, gives realistic binding and saturation of nuclear matter
already at three-loop order. At the same time it gives very good values for the
compression modulus and the asymmetry energy. These are non-trivial
observations, considering that it all works with only one adjustable parameter
which encodes unresolved short-distance dynamics. Of course, questions about
systematic convergence of the in-medium chiral loop expansion still remain and
need to be explored.

In view of the relevant scales in nuclear matter, the importance of
explicit pion degrees of freedom does not at all come unexpected. Many of the
existing models ignore pions, however. They must introduce purely
phenomenological scalar fields with non-linear couplings and freely adjustable
parameters in order to simulate two-pion exchange effects.

Finally, in order to discuss possible contacts with relativistic nuclear mean
field phenomenology, the following working hypothesis suggests itself as a
guide for further steps. Assume that the nuclear matter ground state represents
a "shifted" QCD vacuum characterized by strong vector ($V$) and scalar ($S$)
condensate fields acting on the nucleons, with $V \simeq -S \simeq 0.3 \, GeV$
as suggested e.~g. by in-medium QCD sum rules \cite{13}. Such a scenario would
not produce binding all by itself, but establish strong spin-orbit splitting
and approximate pseudo-spin symmetry \cite{14}. Binding and saturation would
then result from the pionic (chiral) fluctuations around this new vacuum.

\end{document}